\documentstyle[epsfig,floats,aps,prl,twocolumn]{revtex}

\begin{document}

\draft

\title{Quantum Hall Fluctuations and Evidence for Charging in the Quantum 
Hall Effect}

\author{David H. Cobden}
\address{Oersted Laboratory, Niels Bohr Institute, Universitetsparken 5, 
DK-2100 Copenhagen, Denmark}
\author{C. H. W. Barnes and C. J. B. Ford}
\address{Cavendish Laboratory, Madingley Road, Cambridge CB3 OHE, UK}
\date{\today }

\wideabs{
\maketitle
\begin{abstract}
We find that mesoscopic conductance fluctuations in the quantum Hall 
regime in silicon MOSFETs display simple and striking patterns.  The 
fluctuations fall into distinct groups which move along lines parallel to 
loci of integer filling factor in the gate voltage-magnetic field plane.  
Also, a relationship appears between the fluctuations on quantum Hall 
transitions and those found at low densities in zero magnetic field.  
These phenomena are most naturally attributed to charging effects.  We 
argue that they are the first unambiguous manifestation of interactions in 
dc transport in the integer quantum Hall effect.
\end{abstract}
\pacs{PACS numbers: 73.40.Hm, 73.23.Hk, 73.23.-b}
}

The microscopic situation in a two-dimensional (2D) quantum Hall (QH) 
conductor is strongly influenced by interactions, as evinced by the 
observation of fractional resistance plateaus when the disorder is 
weak.  In the integer QH regime, for a sufficiently smooth disorder 
potential, interactions cause the electron liquid to separate into 
metallic (compressible) and insulating (incompressible) regions 
\cite{Efros}, as has recently been confirmed by direct imaging techniques 
\cite{McCormick,vanHaren,Tessmer,Wei,Yacoby}.  
Nevertheless, it may still be argued that all linear-response dc transport 
properties in the integer regime are demonstrated by noninteracting 
models in which electrons penetrate through a disordered potential 
landscape \cite{Huckestein}.  Modifications by interactions appear to be 
subtle or negligible - for example, the localization length exponent is 
unaffected.  This has helped to justify the continuing study 
of noninteracting models.

Here we report that the mesoscopic conductance fluctuations in small Si 
metal-oxide-semiconductor field-effect transistors (MOSFETs) provide 
strong evidence that interactions, in the form of charging effects, can 
have a profound effect on conduction in the integer QH regime.  Previous 
experiments \cite{Timp,Cobden} on these quantum Hall fluctuations (QHFs) have 
focused on their shape and periodicity, for comparison with predictions of the 
peak shapes \cite{Jain} and the conductance distribution \cite{Wang} 
and correlation \cite{Xiong} functions.  Unfortunately, owing to the critical 
dependence on magnetic field and density it is hard to make detailed 
statistical measurements near a QH transition.  Here we concentrate instead 
for the first time on the evolution of the fluctuations 
in the magnetic field-density plane.  Our principal result is as follows: 
the extrema (peaks and dips) in the QHFs fall into groups moving along linear 
trajectories parallel to lines of constant integer filling 
factor $\nu = p$, where $p = 0, 1, 2, ...$.  We
conclude that a realistic picture of the QHE transition must go well 
beyond noninteracting models, to incorporate not only the existence 
of insulating and metallic regions, but also the charging conditions for 
electrons and holes in the metallic regions.

The MOSFETs used have oxide thickness $d_{ox} = 25$ nm and a range of effective 
channel dimensions $L$ and $W$ from 0.4 $\mu$m to 4 $\mu$m \cite{Cobden}. 
Each device is 
approximately a rectangle of disordered two-dimensional electron gas 
(2DEG) with a metallic contact (n+ diffusion) at each end, as sketched in 
the left inset to Figure 1.  The electron density $\rho$ is linear in the 
voltage $V_{g}$ on the metallic polysilicon gate: d$\rho/$d$V_{g} = C/e =$ 
8.6$\times$10$^{11}$~cm$^{-2}$V$^{-1}$, where $C = 
\epsilon_{ox}\epsilon_{0}/d_{ox}$ and $\epsilon_{ox}$ = 3.9 \cite{Ando}.  The 
conductance $G$ was measured using an ac bias of 10$\mu$~V.  The mobility at low 
temperature is of order 0.2~m$^{2}$V$^{-1}$s$^{-1}$, corresponding to a mean 
free path $l \sim$ 20 nm, and the phase coherence length 
was $L_{\phi} \sim$ 0.4$\mu$m \cite{Cobden}
at base temperature ($T \approx$ 100mK) in the dilution refrigerator.

Figure 1 (a) shows the $G-V_{g}$ characteristics of small square device M1.  
As $B$ is increased, quantum Hall plateaus develop, 
which we label by the integers $p =$ 0,1,2,..., where $p = (h/e^2) G$.  
All the fluctuations visible in these characteristics are reproducible.  Similar 
fluctuations are seen in every device, though their amplitude decreases as the 
device area increases.  The fluctuations can be separated into three categories, 
as follows.  The first, seen at low $B$ and $V_{g} \lesssim 2.6$ V (where 
$G \gtrsim e^{2}/h$), are small and have a short period in $V_{g}$.  We refer to 
them as rapid threshold fluctuations (RTFs).  Similar features have earlier been 
attributed to hopping or resonant tunneling \cite{Popovic}.  The second, seen at 
low $B$ and higher $V_{g}$, exhibit the properties of universal conductance 
fluctuations (UCFs), characteristic of the diffusive regime \cite{Skocpol}.  
The third, seen in the transition regions between the QH plateaus, are the 
QHFs.  As $T$ is decreased, the QH transitions narrow while the QHFs sharpen 
and grow, as illustrated in the right inset to Figure 1.  At low $T$ 
the peak-to-dip amplitude near the center of the transition approaches 
$e^{2}/h$, yielding a roughly top-hat-shaped conductance distribution function 
as reported previously \cite{Cobden}.

\begin{figure}
\centerline{\epsfxsize=3.1in\epsfbox{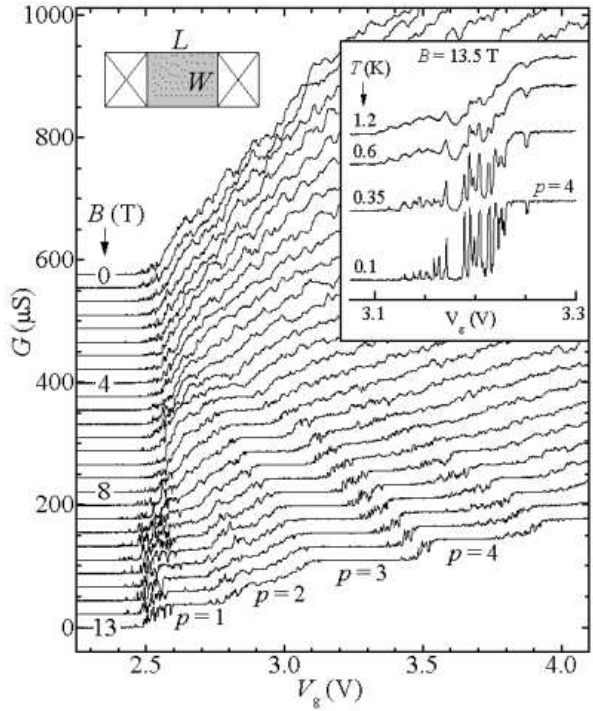}}
\vspace{0.1cm}
\caption{$G-V_{g}$ characteristics of device M1 ($L = W =$ 0.6 $\mu$m) 
in a perpendicular magnetic field $B$.  Left inset: schematic device.  Right 
inset: $T$-dependence of the $p = 3 \rightarrow 4$ transition.}
\label{fig:1}
\end{figure}

To study the QHFs as a joint function of $B$ and $V_{g}$ we make greyscale plots 
of $G$ such as those in Figure 2.  A smooth background is subtracted so that 
the plateaus appear grey, while peaks and dips produce respectively darker 
or lighter regions.  Figure 2 (a) shows data for device M2 at high $B$.  It 
can be seen that the extrema follow long, straight trajectories.  In an 
expanded view of the 3rd ($p = 2 \rightarrow 3$) transition (Figure 2 (b)) it 
can further be seen that the extrema in this region fall into two groups.  
Those in one group move parallel to the dashed line drawn on the $p = 2$ 
plateau, while those in the other move parallel to the dotted line on the 
$p = 3$ plateau.  Each group contains both peaks and dips.  A similar 
pattern is repeated on every transition and in every device, irrespective 
of size and geometry.  Figure 2 (c) shows data for the 3rd transition in a 
larger device, M3.  This plot has been sheared parallel to the $V_{g}$ axis to 
make the center of the transition vertical.  In this case it can be seen 
that the two components of the fluctuations coexist right across the 
transition region.

Combining results for the fluctuations on the first few transitions in 
several devices, we find that, to within a few percent accuracy in all 
cases, the trajectories of the extrema are parametrized by
\begin{equation}
{C\over e}{{\partial V_{g}} \over {\partial B}} = p {e \over h}
\end{equation}
where $p$ is an integer.  Assuming d$\rho/$d$V_{g} = C/e$, this can also be 
written as $\partial \rho / \partial B = pe/h$.  Hence the extrema move parallel 
to lines of integer filling factor, $\nu = (h/e) \rho/B = p$.  Let us now 
discuss the implications of this, our key result.  

\begin{figure}
\centerline{\epsfxsize=3.1in\epsfbox{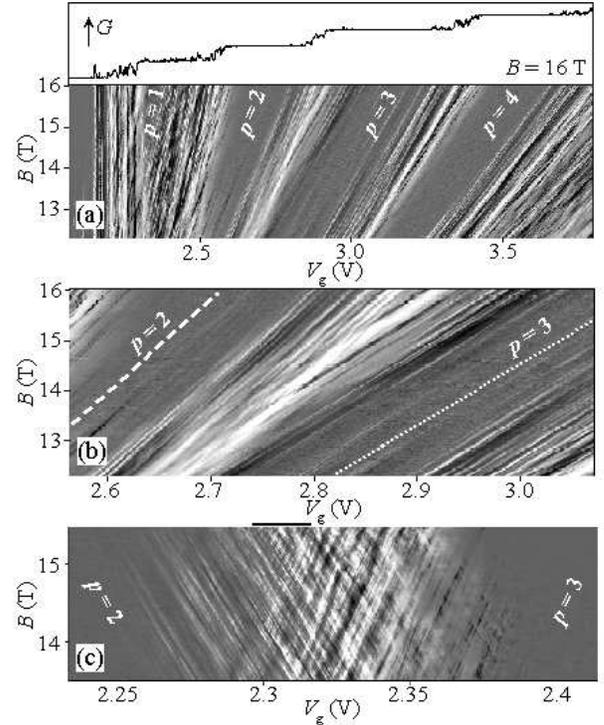}}
\caption{Greyscale plots of conductance with a smooth background subtracted 
(lighter = smaller $G$).  (a) Device M2 ($L = W =$ 0.6 $\mu$m), showing the 
first four QH plateaus, as indicated in the trace above.  (b) Zoom on the third 
transition in (a).  (c) Third transition in device M3 ($L = W = 3.4 \mu$m).  
This plot has been sheared to make the center of the transition vertical.}
\label{fig:2}
\end{figure}

We begin by noting that Eq.~(1) contradicts the predictions of 
noninteracting models, in which extrema are associated with alignment of 
the Fermi level with scattering resonances in particular Landau bands 
\cite{Jain}.  First, such resonances are expected either to follow 
{\it half-integer} filling factors or to show no clear patterns at all 
\cite{simulations}.  Second, owing to the nonuniform density of states at 
high $B$, the lines given by Eq.~(1) in the $V_{g}-B$ plane correspond to 
distorted trajectories in the $E_{F}-B$ plane which do not resemble the 
expected paths of resonances \cite{simulations}.  Third, the 
distance between resonances in different Landau bands should depend on the 
energy gaps between bands.  In MOSFETs the first four transitions are 
caused by spin and valley gaps {\it within the first orbital Landau level} 
\cite{Ando}.  In contrast, we find that Eq.~(1) holds irrespective of spin, 
valley and orbital indices.

However, if we assume instead that interactions are strong enough to make 
the electron density weakly dependent on $B$, we can make more headway with 
Eq.~(1).  Let $\rho(x,y,V_{g},B)$ be the spatial density profile in the 2D 
($x$-$y$) plane of the device, and consider an extremum of index $p$ which 
passes through $(V_{g1},B_{1})$ in the $V_{g}$-$B$ plane.  Eq.~(1), together 
with $d\rho/dV_{g} = C/e$, 
then implies the following: along the path of the extremum, the contour in 
the $x$-$y$ plane defined by $\rho(x,y,V_{g1},0) = p(e/h)B_{1}$ is unchanged.  
This is illustrated in Figure 3, diagrams (a)-(d).  We take a smooth random 
density profile and plot it at $(V_{g1},B_{1})$ in (a) and at 
$(V_{g1}+\Delta V_{g},B_{1}+\Delta B)$ in 
(b), where $(C/e) \Delta V_{g}/ \Delta B = e/h$.  To help interpret these plots 
we show cross-sections through them in (c) and (d).  Because we have chosen 
$\Delta V_{g}/\Delta B$ corresponding to a $p = 1$ trajectory, the 
contour (black) at $\rho = eB/h$ does not change shape between (a) and (b).  
Meanwhile, the contour (white) at $\rho = 2eB/h$ shrinks.

\begin{figure}
\centerline{\epsfxsize=3.1in\epsfbox{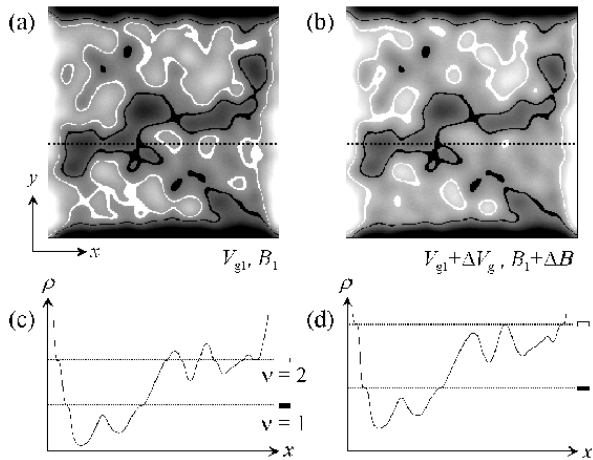}}
\vspace{0.3cm}
\caption{Density $\rho(x,y,V_{g},B)$ at two values of $V_{g}$ and 
$B$.  Diagrams (c) and (d) are cross-sections along the dotted lines through 
contour plots (a) and (b) respectively.  Density contours corresponding to integer 
filling factor are indicated by black ($\nu = 1, \rho = eB/h$) or 
white ($\nu = 2, \rho = 2eB/h$), while other densities are shown in greyscale.}
\label{fig:3}
\end{figure}

In other words, each extremum can be linked to the occurrence of a density matching
an integer $\nu$ at a particular spatial contour in the device.  Now, it has been 
predicted \cite{Efros} and observed \cite{Yacoby,Wei} that for smooth 
disorder the electron liquid becomes incompressible and insulating wherevere the 
local density matches an integer filling factor \cite{finitestrips}.  
Accordingly, an extremum with index $p$ may be associated with a 
particular shape of the $\nu = p$ incompressible strip.
This leads naturally to an explanation for Eq.~(1) \cite{vanHouten,Chklovskii}.  
Whevener the $\nu = p$ incompressible strip completely surrounds a metallic region, 
as occurs several times in Figure 3, the charge on the metallic puddle 
should be quantized.  The charging condition is determined 
mainly by the capacitance to the nearby gate, and therefore by the puddle's area.  
Its charge state will thus not change as long as its shape is maintained, which is 
the case if $V_{g}$ and $B$ are varied according to Eq.~(1).

Let us now consider other properties of the QHFs to see if they support 
this charging picture.  The peak separation $\Delta V_{g}$ is always 
$\gtrsim 1$mV at base 
$T$.  The conductance at an extremum, followed along its trajectory, varies 
in amplitude on a scale of 1 to 2 T (see e.g. Figure 2 (b)).  Also, the 
QHFs sometimes exhibit periodicity, such as may be discerned in the region 
indicated by a horizontal bar above Figure 2 (c).  In all these 
properties, as well as in their qualitative $T$ dependence (Figure 1 inset), 
they resemble the RTFs.  There is however a more direct link between the 
QHFs and RTFs, which is illustrated in Figure 4.  Figure 4 (a) shows the 
first two transitions in device M4.  As on higher transitions, the extrema 
on the first transition ($p = 0 \rightarrow 1$) fall into two groups.  One group 
is associated with the $p = 1$ plateau.  The other is independent of $B$, i.e., 
it corresponds to $p = 0$ in Eq.~(1).  Figure 4 (b) shows data for device M5 
over a wide range of $B$.  The RTFs at low $B$ can be seen to evolve into the 
$p = 0$ QHFs at high $B$, with no qualitative change in their properties.  The 
RTFs may effectively be identified with the $p = 0$ QHFs.

\begin{figure}
\centerline{\epsfxsize=3.1in\epsfbox{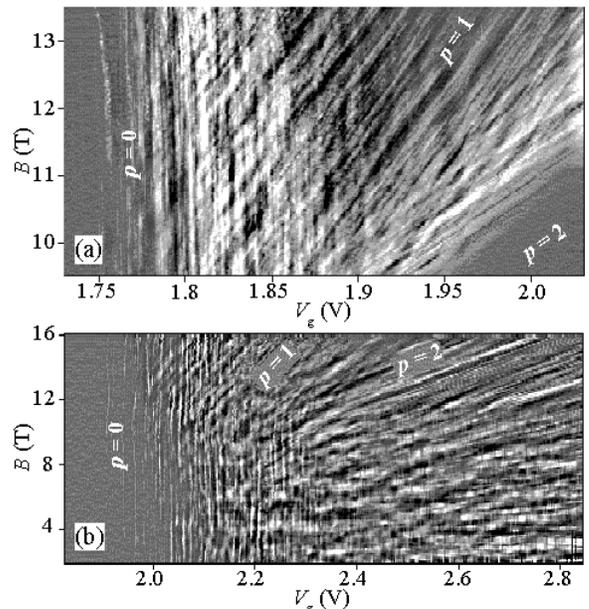}}
\caption{Greyscale plots showing fluctuations near threshold.  (a) First 
two QH transitions, device M4 ($L = 1.4  \mu m, W = 3.4 \mu m$).  (b) Wide range 
of B, device M5 $(L = 0.4 \mu m, W = 0.6 \mu m$).}
\label{fig:4}
\end{figure}

In MOSFETs with dimensions smaller than about 0.2 $\mu$m, the RTFs are often highly 
periodic \cite{vanHouten,ScottThomas}.  This is a signature of Coulomb blockade. 
The following scenario therefore seems likely: near threshold, conduction is 
limited by bottlenecks separating puddles of electrons \cite{vanHouten}.  In 
devices not much larger than the typical puddle size, one puddle often dominates 
the conductance, resulting in Coulomb blockade oscillations.  Consistent 
with this, the minimum peak spacing of $\Delta V_{g} \gtrsim 1$mV in our devices 
corresponds to adding one electron to a maximum puddle area of 
$e/(C\Delta V_{g}) \lesssim$ (0.3 $\mu$m)$^{2}$, 
which is one quarter of the area of our smallest device and 
contains of order a thousand electrons.  The insensitivity of the 
RTFs (the $p = 0$ lines) to $B$ reflects the $B$-independent charging spectrum 
of these large puddles.  The aperiodicity in larger devices 
results from charging of many puddles simultaneously.

The above evidence that the RTFs are strongly influenced by charging, 
together with the link demonstrated between the QHFs and the RTFs, 
firmly supports the inference from Eq.~(1) that the QHFs are 
dominated by charging effects.
We emphasize however that we do not propose a detailed model for the 
QHFs.  The picture of smooth disorder illustrated in Figure 3 may not 
accurately describe MOSFETs, where the disorder is produced by fixed 
charges near the interface \cite{Ando,Hori}.  In its defence we note that the 
effective potential may be softened by electrons bound to the fixed 
charges \cite{threshold}, and that the results of arguments based on a smooth 
potential often have a surprisingly general validity.  Nevertheless, other 
difficult questions also arise in this model.  Particularly challenging is 
the coexistence of $p = i$ and $p = i+1$ extrema on the $i$th transition.  This 
requires density variations larger than $eB/h$, to allow
different incompressible strips to be present simultaneously, as in Figure 
3.  (For lower disorder the situation would presumably be different.)  It 
also requires the simultaneous presence of $p = i$ peaks and $p = i+1$ dips, 
which is hard to realize if these features are Coulomb blockade 
oscillations and if wide plateaus (where one insulating strip percolates) 
are to be reproduced.  Another interesting issue is that the unity 
amplitude of the QHFs in the smallest devices \cite{Cobden} seems to require 
phase-coherence over the entire system, even as charging takes place.

Finally, we note that our results may have a bearing on the $B = 0$ 
metal-insulator transition in larger MOSFETs \cite{Kravchenko95} and its relation to 
the integer QH effect \cite{Kravchenko98}.  Further, they support the findings of 
other recent experiments that single-particle scaling theory may not always be
appropriate for describing QH transitions \cite{Shahar}.

In summary, our analysis of mesoscopic fluctuations has led us to conclude 
that interactions, in the form of charging effects, have a profound 
influence on transport in the integer quantum Hall effect.

We thank Y. Oowaki of Toshiba for supplying the devices,  J. T. Nicholls for 
invaluable experimental assistance, and amongst others J. Chalker, D. 
Chklovskii, N. Cooper, V. Fal'ko, L. Glazman, E. Kogan, D.-H. Lee, D. Maslov, 
P. L. McEuen, M. Pepper, M. E. Raikh and Z. Wang for helpful discussions.  
This work was supported by the UK EPSRC and by an EU TMR grant.


\begin{references}

\bibitem{Efros} A. L. Efros, Phys. Rev. B {\bf 45} 11354 (1992); 
D.B. Chklovskii; B. I. Shlovskii, and L. L. Glazman, Phys. Rev. B {\bf 46}, 4026 (1992); N.R. Cooper and J. T. Chalker, Phys. Rev. B {\bf 48}, 4530 (1993); 
V. Tsemekhman {\it et al}., Phys. Rev. B {\bf 35}, R10201 (1997).

\bibitem{McCormick} K. L. McCormick {\it et al}., to appear in Phys. Rev. B.

\bibitem{vanHaren} R. J. F. van Haren, F. A. P. Blom, and J. H. Wolter, Phys. Rev. Lett. 
{\bf 74}, 1198 (1995).

\bibitem{Tessmer} S. H. Tessmer {\it et al}., Nature {\bf 392}, 51 (1998).

\bibitem{Wei}Y. Y. Wei {\it et al}., Phys. Rev. Lett. {\bf 81}, 1674 (1998).

\bibitem{Yacoby} A. Yacoby {\it et al}., (not yet in print)

\bibitem{Huckestein} B. Huckestein, Rev. Mod. Phys. {\bf 67}, 357 (1995).

\bibitem{Timp} G. Timp {\it et al}., Phys. Rev. Lett. {\bf 59}, 732 (1987);C. J. B. Ford {\it et al}., Phys. Rev. B {\bf 38}, 8518 (1988); 
J. A. Simmons {\it et al}., Phys. Rev. B {\bf 44}, 12933 (1991); 
P. C. Main {\it et al}., Phys. Rev. B {\bf 50}, 4450 (1994); 
A. A. Bykov et al., Phys. Rev. B {\bf 54}, 4464 (1996).

\bibitem{Cobden} D. H. Cobden and E. Kogan, Phys. Rev. B {\bf 54}, R17316 (1996).

\bibitem{Jain} J. K. Jain and S. A. Kivelson, Phys. Rev. Lett. {\bf 
60}, 1542 (1988).

\bibitem{Wang} Z. Wang, B. Jovanovic, and D.-H. Lee, Phys. Rev. Lett. {\bf 77}, 4426 (1996); A. G. Galstyan and M. E. Raikh, Phys. Rev. B {\bf 56}, 1422 (1997); 
S. Cho and M. P. A. Fisher, Phys. Rev. B {\bf 55}, 1637 (1997).

\bibitem{Xiong} S. Xiong and A. D. Stone, Phys. Rev. Lett. {\bf 68}, 3757 (1992); 
D. L. Maslov and D. Loss, Phys. Rev. Lett. {\bf 71}, 4222 (1993); 
D. E. Khmelnitskii and M. Yosefin, Surf. Sci. {\bf 305}, 507 (1994); 
B.Jovanovic and Z. Wang, Phys. Rev. Lett. {\bf 81}, 2767 (1998).

\bibitem{Ando} T. Ando, A. B. Fowler, and F. Stern, Rev. Mod. Phys. {\bf 54} 437 (1982)

\bibitem{Popovic} D. Popovic, S. Washburn, and A. B. Fowler, Internat. J. Mod. Phys. B, 
{\bf 8}, 809 (1994) and references therein.

\bibitem{Skocpol} W. J. Skocpol {\it et al}., Phys. Rev. Lett. {\bf 56}, 2865 (1986); Mesoscopic Phenomena in Solids, eds. B. L. Altshuler, P. A. Lee and R. A. Webb 
(North-Holland, Amsterdam, 1991).

\bibitem{simulations} In single-particle simulations for a variety of disorder potentials 
we found that trajectories of resonances never form sets of straight or parallel lines.

\bibitem{finitestrips} The true finite extent of the incompressible regions is indicated by the broadening of contours in Fig. 3.

\bibitem{vanHouten} H. van Houten, C. W. J. Beenakker, and A. A. M. Staring, 
in Single Charge Tunneling, eds. H. Grabert and M. H. Devoret, NATO ASI 
Series B vol. {\bf 294} (Plenum, New York, 1991).

\bibitem{Chklovskii} D. B. Chklovskii, preprint, cond-mat/9609023;
B. W. Alphenaar {\it et al}., Phys. Rev. B {\bf 46}, 7236 (1992).

\bibitem{ScottThomas} We observed periodic oscillations in 100 nm-wide 
devices, as did J. H. F. Scott-Thomas {\it et al}., Phys. Rev. 
Lett. {\bf 62}, 583 (1989) and M. G. Peters, PhD thesis (U. Utrecht, 1997).

\bibitem{Hori} See, e.g., T. Hori, Gate Dielectrics and MOS ULSIs (Springer-Verlag, 
Berlin, 1997).

\bibitem{threshold} At threshold the electron density 
is around 10$^{12}$cm$^{-2}$, judged from Shubnikov-de Haas measurements.

\bibitem{Kravchenko95} S. V. Kravchenko {\it et al}., Phys. Rev. B {\bf 51}, 7038 (1995); D. Popovic, A. B. Fowler, and S. Washburn, Phys. Rev. Lett. {\bf 79}, 1543 (1997).

\bibitem{Kravchenko98} S. Kravchenko {\it et al}., preprint, cond-mat/9812389.

\bibitem{Shahar} D. Shahar {\it et al}., Solid State Commun. {\bf 107}, 19 (1998); 
N. Q. Balaban, U. Meirav, and I. Bar-Joseph, Phys. Rev. Lett. {\bf 81}, 4967 (1998).

\end{references}
\end{document}